\documentclass{article}

\newif\ifclassllncs{} 
\newif\ifclassbeamer{} 
\newif\ifclassmimosis{}
\newif\ifpackagebabel{} 
\newif\ifclasssnjnl{}
\makeatletter%
\@ifclassloaded{llncs}%
  {\classllncstrue}%
  {\classllncsfalse}%
\@ifclassloaded{beamer}%
{\classbeamertrue}%
{\classbeamerfalse}%
\@ifclassloaded{mimosis}%
{\classmimosistrue}%
{\classmimosisfalse}%
\@ifpackageloaded{babel}%
{\packagebabeltrue}%
{\packagebabelfalse}%
\@ifclassloaded{sn-jnl}%
{\classsnjnltrue}%
{\classsnjnlfalse}%
\makeatother%

\usepackage[T1]{fontenc}
\usepackage[utf8]{inputenc}
\usepackage{lmodern}
\usepackage[english]{babel}

\usepackage{graphics}
\usepackage{setspace}
\usepackage{xcolor}
\usepackage{xparse}

\usepackage{metalogo}

\usepackage{etoolbox}

\usepackage{bookmark}

\ifclassllncs{}
    \usepackage[caption=false]{subfig}
\else 
    \ifclassmimosis{}
    \else
        \usepackage[format=hang]{caption} 
        \usepackage{subfig}
    \fi
\fi

\ifclassllncs{}
     
\fi

\usepackage{amsmath}
\usepackage{amsthm} 
\usepackage{amssymb}
\usepackage{pifont}

\usepackage{thmtools,thm-restate}

\usepackage{fontawesome}

\usepackage{mathtools}
\usepackage{interval}
\intervalconfig{soft open fences}

\usepackage{hyperref}

\ifclassbeamer{}
\else
    \usepackage{cleveref}
    \usepackage{footnote}
    \makesavenoteenv{tabular}
    \makesavenoteenv{figure}
    \makesavenoteenv{table}
\fi

\usepackage{hypcap}

\usepackage{tabularx}
\usepackage{booktabs}

\usepackage{hyphenat}

\usepackage{msc}

\usepackage{ifthen}

\usepackage{csquotes}

\ifclassmimosis{}
    
\fi
\usepackage[n, advantage, operators, sets, adversary, landau, probability, notions, logic, ff, mm, primitives, events, complexity, oracles, asymptotics, keys]{cryptocode}

\usepackage{tikz} 
\usetikzlibrary{arrows.meta,backgrounds,positioning,arrows,bending,shapes.multipart,decorations.pathreplacing,fit,shapes.geometric}

\ifpackagebabel{}
    \usetikzlibrary{babel}
\fi

\usepackage{multirow} 

\usepackage{tikzducks}
\usepackage{tikzpeople}

\usepackage{ragged2e}

\newif\ifhidecomment{}
\hidecommenttrue{}
\hidecommentfalse{} 

\ifclassbeamer{}
    \usepackage[duration=30]{pdfpc}
\else

\fi

\newif\ifshowauthor{}
\showauthortrue{}

\ifclassllncs{}
\else
    \ifclasssnjnl{}
        \theoremstyle{thmstyleone}%
        \newtheorem{theorem}{Theorem}%
        \newtheorem{proposition}[theorem]{Proposition}%
        \newtheorem{lemma}[theorem]{Lemma}

        \theoremstyle{thmstyletwo}%
        \newtheorem{example}{Example}%
        \newtheorem{remark}{Remark}%

        \theoremstyle{thmstylethree}%
        \newtheorem{definition}{Definition}%
    \else
        \ifclassbeamer{}
        \else
            \theoremstyle{definition}

        \fi
    \fi
\fi

\definecolor{amaranthred}{rgb}{0.83,0.13,0.18}
\definecolor{forestgreenweb}{rgb}{0.13,0.55,0.13}
\newcommand{\cmark}{\textcolor{forestgreenweb}{\ding{51}}}
\newcommand{\xmark}{\textcolor{amaranthred}{\ding{55}}}

\usepackage{orcidlink}

\newcommand{\qmark}{\textcolor{orange}{\ding{58}}}

\newcommand{\poserandom}{\textbf{PoSE\textsubscript{random}}}
\newcommand{\posegraph}{\textbf{PoSE\textsubscript{graph}}}
\newcommand{\poselight}{\textbf{PoSE\textsubscript{light}}}
\newcommand{\dziembowski}{\textbf{DFKP}}
\newcommand{\karvelas}{\textbf{KK}}
\newcommand{\perito}{\textbf{PT}}
\newcommand{\karame}{\textbf{KL}}

\newcommand{\bigOlog}[1]{\ensuremath{\tilde{\mathcal{O}}(#1)}}

\newcommand{\first}{(i)} 
\newcommand{\second}{(ii)} 
\newcommand{\third}{(iii)} 

\def\securityprop{
	\begin{table*}[!ht]
		\footnotesize
		\centering
		\begin{tabular}{lccccccc}
			& \textbf{Proof} & \textbf{Prob.} & \textbf{No-Isolation} &  
			\textbf{Erasure} & \textbf{Time} & \textbf{Comm.} \\
		   \dziembowski{}~\cite{Dziembowski2011} 
		   & \cmark{} &
		   \xmark{} & \xmark{} & 
		   \( 1 \) & \(\bigO{n^2}\) & \(\bigO{1}\) \\
		   \karame{}~\cite{Karame2015} 
		   & \xmark{} & \xmark{} & \xmark{} & 
		\( 1 \) & \(\bigO{n}\) & \(\bigO{n}\) \\
		   \karvelas{}~\cite{Karvelas2014} 
		   & \cmark{} & \xmark{} & \xmark{} & 
		   \( \frac{1}{32} \) & \(\bigOlog{n}\) & \(\bigO{1}\) \\
		   \perito{}~\cite{Perito2010} 
		   & \xmark{} & \xmark{} & \xmark{} & 
		   \( 1 \) & \(\bigO{n}\) & \(\bigO{n}\) \\
		   \posegraph~\cite{Bursuc2024} 
		   & \cmark{} & \cmark{} & \cmark{} & 
		\( f(r, n, p) \) & \(\bigOlog{n+r}\) & \(\bigO{r}\) \\
		   \poselight~\cite{Bursuc2024a} 
		   & \cmark{} & \cmark{} & \cmark{} & 
		\( f(r, n, p) \) & \(\bigO{n+r}\) & \(\bigO{r}\) \\
		   \poserandom~\cite{Bursuc2024} 
		   & \cmark{} & \cmark{} & \cmark{} & 
		\( f(r, n, p) \) & \(\bigO{n+r}\) & \(\bigO{n+r}\) \\
		   \textbf{TR}~\cite{Trujillo-Rasua2019} 
		   & \cmark{} &
		   \xmark{} & \cmark{} & 
		   \xmark{} & \(\bigO{n + r}\) & \(\bigO{r}\) \\
		\textbf{SPEED}~\cite{Ammar2018} 
		& \xmark{} & \xmark{} & \cmark{} & 
		1 & \(\bigO{n}\) & \(\bigO{1}\) \\
		\end{tabular}
	\caption{\label{tab:secprop}Characteristics of the implemented
  memory-erasure protocols. The asymptotic bounds for the execution time and
  communication complexity are given in terms of the memory size (denoted \( n \)) and the number of round-trip
  measurements (denoted \( r \)). The erasure guarantees of the three
  protocols in~\cite{Bursuc2024,Bursuc2024a}  depend on \( r \), \( n \) and a bound \( p \)
  on the desired probability of success of the attacker. As an example,
  given \( r = 71 \), \( n = 8\)KB and \( p = 10^{-3} \), \( f(r, n, p) = 0.9 \).
   }
\end{table*}
}

\def\securityproptime{
	\begin{table*}[!ht]
		\centering
		\scriptsize
		\begin{tabular}{lccccccccccc}
			& \textbf{Proof} & \textbf{Prob.} & \textbf{No-Isol.} &  
			\multicolumn{2}{c}{\textbf{Erasure}}  & \multicolumn{6}{c}{\textbf{Total Time}} \\
			&  &  &  &  &  & \multicolumn{2}{c}{\textbf{F5529\-2KB}} & \multicolumn{2}{c}{\textbf{FR5994\-2KB}} & \multicolumn{2}{c}{\textbf{CC2652\-8KB}} \\
		   \dziembowski{}~\cite{Dziembowski2011} & \cmark{} & \xmark{} &
		   \xmark{} & 
		   \( 1 \) & \cmark & 10.0 & \qmark & 3.2 & \cmark & 1.9 & \cmark \\
		   \karame{}~\cite{Karame2015} & \xmark{} & \xmark{} & \xmark{} & 
		\( 1 \) & \cmark & 2.9 & \cmark & 2.4 & \cmark & 23.3 & \xmark \\
		   \karvelas{}~\cite{Karvelas2014} & \cmark{} & \xmark{} & \xmark{} &
		   \( \frac{1}{32} \) & \xmark & 16.7 & \qmark & 5.1 & \cmark & 1.3 & \cmark \\
		   \perito{}~\cite{Perito2010} & \xmark{} & \xmark{} & \xmark{} & 
		   \( 1 \) & \cmark & 2.9 & \cmark & 2.4 & \cmark & 23.2 & \xmark \\
		   \posegraph~\cite{Bursuc2024} & \cmark{} & \cmark{} & \cmark{} & 
		\( 0.9 \) & \qmark & 31.1 & \xmark & 13.3 & \xmark & 8.7 & \qmark \\
		   \poselight~\cite{Bursuc2024a} & \cmark{} & \cmark{} & \cmark{} & 
		\( 0.9 \) & \qmark & 18.3 & \qmark & 9.3 & \qmark & 7.2 & \qmark \\
		   \poserandom~\cite{Bursuc2024} & \cmark{} & \cmark{} & \cmark{} & 
		\( 0.9 \) & \qmark & 6.8 & \cmark & 6.8 & \qmark & 29.7 & \xmark \\
	\end{tabular}
	\caption{\label{tab:secproptime}Summary of results, contrasting performance with security. The last three columns show the total time
   on each device, when using the highest possible memory to erase (in
   parentheses) and the hash function leading to the lowest execution time. }
\end{table*}
}

\def\devprop{
	\begin{table*}[!ht]
		\centering
		\small
		\begin{tabular}{lccccccc}
			Device & Memory & Clock & Crypto & BT & Arch & MCU & IoT \\
			F5529 & (128 + 10) KB & 25MHz & --- & 2 & RISC-16 & MSP430 & Class 1
			\\
		   FR5994 & (256 + 8) KB & 16MHz &  AES & 2 & RISC-16 & MSP430 & Class
		   1/2 \\
		   CC2652 & (352 + 88) KB & 48MHz & AES,SHA2 & 5.2 & RISC-32 & Arm
		   Cortex-M & Class 2 \\
		\end{tabular}
	\caption{\label{tab:devprop}Microcontroller characteristics}
\end{table*}
}

\def\resultsummary{
	\begin{table*}[!ht]
		\centering
		\footnotesize
	  \begin{tabular}{|c|c|c|c|c|c|c|c|}
		\cline{5-8}
		\multicolumn{4}{c|}{} & \multicolumn{4}{c|}{Network cost} \\
		  \cline{5-8}
		  \multicolumn{4}{c|}{} &\multicolumn{2}{c|}{High} &
		  \multicolumn{2}{c|}{Low} \\
		\cline{5-8}
		\multicolumn{4}{c|}{} & \multicolumn{2}{c|}{Clock speed} &
		\multicolumn{2}{c|}{Clock speed} \\
		\cline{5-8}
		\multicolumn{4}{c|}{} & High &  Low & High & Low \\
		\hline
		\multirow{4}{*}{Memory size} & \multirow{2}{*}{Large} &
		\multirow{2}{*}{Security} & High & \poselight{} & \poselight{} &
		\poserandom{} & \poserandom{} \\
		\cline{4-8}
		&  & & Low & \karvelas{} & \karvelas{} & \perito{} & \perito{} \\
		\cline{2-8}
		& \multirow{2}{*}{Small} & \multirow{2}{*}{Security} & High &
		\dziembowski{} & \dziembowski{} & \dziembowski{} & \poserandom{} \\
		\cline{4-8}
		&  & & Low & \dziembowski{} & \dziembowski{} & \karvelas{} & \perito{}
		\\
		\hline
	  \end{tabular}
	  \caption{\label{tab:resultsummary}Summary of results. For each combination
	  of Network cost (high, low), Clock speed (high, low), Memory size (large,
	  small) and Security (high, low) the most performant protocol is shown.}
	  \end{table*}
}

\def\totalTimeDevCCs{
	\begin{figure*}[!ht]
		\centering
		\includegraphics[width=0.75\textwidth]{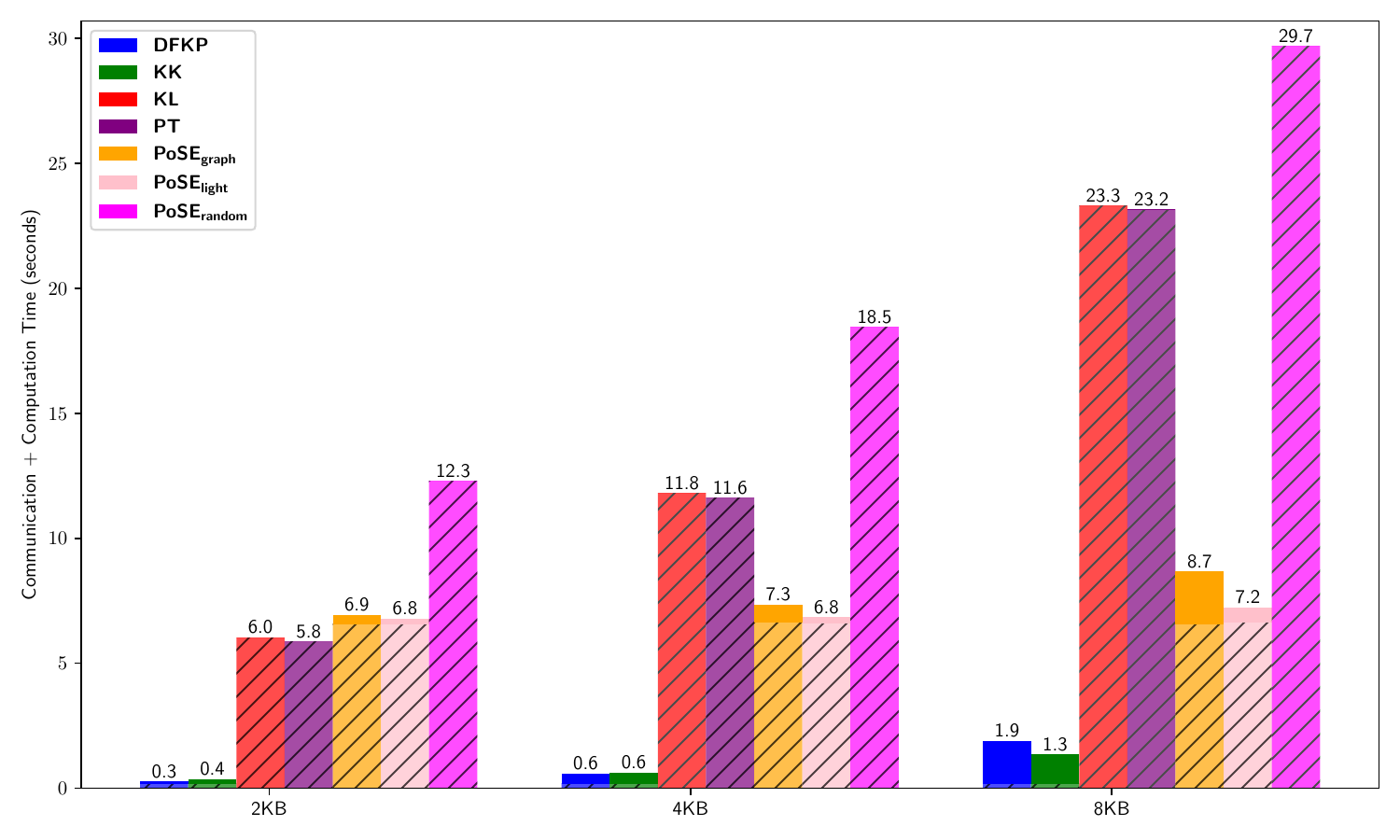}
	\caption{\label{fig:totalTimeDevCCs}Total execution time in seconds on
	device CC2652, partitioned by communication time (striped) and
	computation time, using the best hash function.}
\end{figure*}
}

\def\totalTimeDevsVI{
	\begin{figure*}[!ht]
		\centering
		\includegraphics[width=0.75\textwidth]{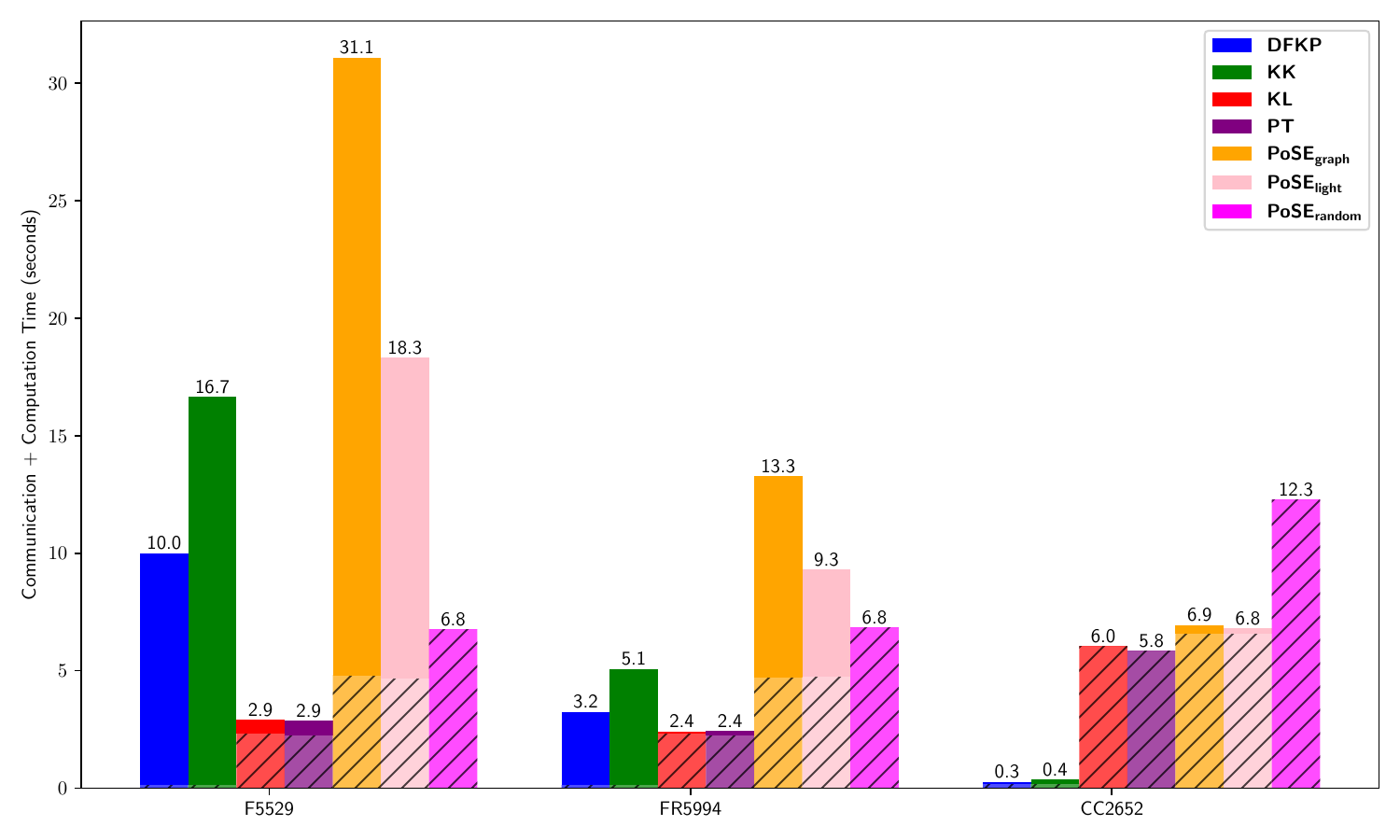}
	\caption{\label{fig:totalTimeDevsVI}Total execution time in seconds,
	partitioned by communication time (striped) and computation time, using
	the best hash function for each device.}
\end{figure*}
}

\usepackage{authblk}       
\usepackage{orcidlink}     
\usepackage{hyperref}      
\usepackage{cleveref}      

\title{Empirical Evaluation of Memory-Erasure Protocols\thanks{This is a preprint version of the article appearing at SECRYPT 2025 with DOI: \texttt{10.5220/0013554800003979}}}

\author[1]{Reynaldo Gil-Pons\orcidlink{0000-0003-1804-3319}}
\author[1]{Sjouke Mauw\orcidlink{0000-0002-2818-4433}}
\author[2]{Rolando Trujillo-Rasua\orcidlink{0000-0002-8714-4626}}

\affil[1]{University of Luxembourg, 6 Avenue de la Fonte, Belval, Luxembourg}
\affil[2]{Universitat Rovira i Virgili, 26 Avinguda dels Països Catalans, Tarragona, Spain}

\date{} 

\begin{document}

\maketitle

\begin{abstract}
Software-based memory-erasure protocols are two-party communication protocols where a verifier instructs a computational device to erase its memory and send a proof of erasure. They aim at guaranteeing that low-cost IoT devices are free of malware by putting them back into a safe state without requiring secure hardware or physical manipulation of the device. Several software-based memory-erasure protocols have been introduced and theoretically analysed. Yet, many of them have not been tested for their feasibility, performance and security on real devices, which hinders their industry adoption. This article reports on the first empirical analysis of software-based memory-erasure protocols with respect to their security, erasure guarantees, and performance. The experimental setup consists of 3 modern IoT devices with different computational capabilities, 7 protocols, 6 hash-function implementations, and various performance and security criteria. Our results indicate that existing software-based memory-erasure protocols are feasible, although slow devices may take several seconds to erase their memory and generate a proof of erasure. We found that no protocol dominates across all empirical settings, defined by the computational power and memory size of the device, the network speed, and the required level of security. Interestingly, network speed and hidden constants within the protocol specification played a more prominent role in the performance of these protocols than anticipated based on the related literature. We provide an evaluation framework that, given a desired level of security, determines which protocols offer the best trade-off between performance and erasure guarantees.
\end{abstract}

\vspace{1em}

\section{\uppercase{Introduction}}

The low-cost characteristics of some classes of IoT (Internet of Things) devices
and the ecosystem in which they operate make them particularly vulnerable to
malware infection and an attractive target for cyberattackers. Defenders, in a
context of limited computational resources, cannot rely on health-monitoring
software installed on the IoT device to fight malware. Instead, defenders ought
to resort to security services provided by an external agent that, by
interacting with the IoT device and observing its behaviour, can conclude
whether the device is malware-free. 

There exist two fundamental security services to ensure that
computationally-constrained devices are malware-free: memory attestation and
memory erasure. The former starts from an expectation on the contents of a
device's memory and the assumption that its contents are malware-free, followed
by a mechanism to verify the integrity of the device's memory. That is, memory
attestation checks whether a device's memory is in a \emph{known} safe state.
Memory erasure, instead, replaces the contents of a device's memory with random
data, effectively removing all information, including malware, from the device.
Both memory attestation and memory erasure have their use-cases. Our focus in
this article is on memory erasure, which allows defenders to remove malware
without making assumptions on the contents of the device's memory. This service
is particularly useful for updating the memory of a device with new software
securely, allowing the device to be redeployed elsewhere.

Memory-erasure mechanisms that depend on having physical access to the IoT device or using dedicated secure hardware on the
IoT device, such as a Trusted Platform Module,  
have the potential to offer high security assurances. However, the
first requirement makes memory erasure unscalable and
time-consuming, while the second requirement makes memory erasure expensive or infeasible in low-cost devices. Mechanisms requiring none of those features, which are, hence,
suitable for low-cost IoT devices, are called \emph{software-based
memory-erasure} protocols. In theory, these  protocols have the advantage of
being able to operate on low-cost, even legacy, devices. In practice, they have
not been thoroughly evaluated for their feasibility on off-the-shelf IoT
devices. 

Before introducing software-based memory erasure, it is important to make a
distinction between \emph{permanent memory erasure} (a.k.a.\ data destruction)
and the type of memory erasure we are referring to. Permanent memory erasure
requires the erasure process to be irreversible against advanced forensics
techniques~\cite{Reardon2013}, including those that rely on physical access to
the device. Irreversibility, however, is not necessarily required  for malware
removal, hence the literature on memory
erasure~\cite{Perito2010,Karvelas2014,Karame2015} where our work fits in does
not aim at irreversibility.

Software-based memory erasure is a two-party communication
protocol executed between a verifier, acting as a powerful computational device,
and a prover, acting as a devise with limited computational resources. The role
of the verifier is to instruct the prover to fill its memory with random data
and verify the erasure proof generated by the prover.  Even though several
software-based memory-erasure protocols have been proposed, they have never been
compared within the same experimental setting. Moreover, those that have been
implemented and tested on real-world
devices~\cite{Perito2010,Karame2015,Ammar2018}, have not made their source-code
publicly available for further scrutiny and analysis. Hence, it is still an open
question how existing software-based memory-erasure protocols perform on
low-cost IoT devices and how they compare in terms of computational and
communication complexity, erasure guarantees, and security.  
This article provides an answer, which we argue is a necessary step towards the
adoption of software-based memory-erasure protocols by the IoT industry. 

\noindent \emph{A brief discussion on related work.} The special characteristics
of constrained IoT devices have pushed practitioners to find the most performant
and efficient algorithms for each task. Ultimately, testing the behaviour of an
algorithm requires deploying them on real devices and conducting performance
evaluations. This has been done recently for lightweight hash
functions~\cite{Rao2019}, cryptographic algorithms~\cite{Silva2024}, and data
protection mechanisms \cite{Lachner2019}. However, as pointed out by recent
surveys on the topic~\cite{Banks2021,Kuang2022}, such level of scrutiny has not
yet been achieved for memory erasure and memory attestation. That is not to say
that these protocols have never been compared with each other. For example, Aman
et al.~\cite{Aman2020} compare their memory-attestation protocol against three
alternatives from the literature. In the case of memory erasure, Karame and
Li~\cite{Karame2015} and Perito and Tsudik~\cite{Perito2010} evaluate the
performance of their protocols on off-the-shelf IoT devices, but do not directly
compare them, nor implement other proposals. A common issue of these examples is
the lack of available open-source implementations, making it hard to reproduce
their evaluation and to comprehensively compare  existing proposals within a
common empirical setting.

\noindent \emph{Contributions.}
This article provides the first comparison and evaluation of software-based
memory-erasure protocols by directly observing and analysing their performance
in a real-world experimental setting. We claim this to be a necessary step
towards their adoption by industry and their deployment in the real-world.  
Crucially, we aim at answering the following questions about existing
software-based memory-erasure protocols: 

\begin{enumerate}
    \item Can they be implemented in low-cost IoT devices? If so, what is their
    memory footprint and execution time? 
    \item How much is their execution time affected by the computational power
    of the device, the size of the memory to erase, the implementation of the
    underlying hash function, and the speed of the communication channel? 
    \item Is there a dominant protocol in terms of performance and security,
    i.e.\ a protocol that performs better than the others in all settings? If
    not, what are the best trade-offs?     
\end{enumerate}

Our experimental setting consists of three off-the-shelf IoT devices from the
microcontroller unit family MCU, namely F5529, FR5994 and CC2652. The first one
does not have cryptographic accelerator support, the second one comes with AES
built-in, and the third one supports AES and SHA256. Aiming at a more
comprehensive evaluation, we provided those three devices with software-based
implementations of various prominent hash functions.

We implemented seven software-based memory-erasure protocols and evaluated them
in terms of performance, security and erasure guarantees. In particular, we
measured the overall time of the protocol execution in relation to the speed of
the communication channel, the device's computational characteristics and the
choice of the hash function implementation. We placed the obtained performance
values alongside other relevant protocol features, such as their security and
erasure guarantees, with the goal of determining the protocols offering optimal
trade-offs. Specifically, given a desired security level, we identify which
protocols strike the best balance between performance and erasure assurances. 

\noindent \emph{Structure of the article.} 
The next section (\Cref{sec-survey}) describes existing software-based
memory-erasure protocols, highlights their most relevant features, and provides
a comparison and analysis based on those features. The remainder of this article
is dedicated to extending that analysis with empirical data on the execution
time of memory-erasure protocols, with the goal of determining their feasibility
and providing a more detailed comparison. \Cref{sec-setup} provides our
experimental setup, \Cref{sec-results} reports on the results obtained after
running the experiments, and  \Cref{sec-discussion} discusses the results.
Concluding remarks are given in \Cref{sec-conclusion}.     

\section{\uppercase{Software-based memory erasure: a survey}}\label{sec-survey}

The earliest software-based protocol for secure memory erasure was introduced by
Perito and Tsudik in 2010~\cite{Perito2010}. At the start of the protocol, the
verifier sends a fresh random value of the same size as the prover's memory. The
prover computes the HMAC of this value, using as key the last bits received. The
security proof of this protocol relies on the fact that, to compute the HMAC on
a random value, the prover needs to first receive the key. The proof of security
is informal, though, and does not specify which assumptions the HMAC function
must fulfil to make the protocol secure. This protocol, in addition, has the
drawback of sending over the network a message as large as the prover's memory. 
An improvement upon Perito and Tsudik's protocol in terms of efficiency was
later proposed by  Karame and Li~\cite{Karame2015}.

To reduce the size of the random value sent over the network by the verifier,  
Dziembowski, Kazana and Wichs~\cite{Dziembowski2011} introduced a software-based
memory-erasure protocol that, given a nonce of standard size, asks the prover to
store in memory a random labelling function.  
Ideally, this step requires the use of the entire memory of the prover. 
The verifier then challenges the prover to quickly give the output of the
labelling function on a number of inputs. Dziembowski, Kazana and Wichs provide
an elegant security proof based on a reduction to a combinatorial game known as
\emph{pebbling}, placing the first stone for the study of memory-erasure
protocols whose security depends on the structure of a graph-based labelling
function. The drawback of their protocol is that the computational complexity of
building the labelling function is quadratic in terms of the memory size of the
prover, which might quickly become inefficient as the prover's memory increases.
This computational complexity was reduced by Karvelas and
Kiayias~\cite{Karvelas2014}, using a labelling function whose underlying graph
has quasilinear size in terms of the memory size of the prover. A major
limitation of this design is that it can only provably erase \( \frac{1}{32} \)
of the prover's memory. This means that Karvelas and Kiayias's protocol offers a
lower erasure guarantee than the protocols we have reviewed so far.  The
work in~\cite{Karvelas2014} also introduces an erasure protocol that is not
based on graphs, but on hard-to-invert hash functions. The security of this
protocol is proven assuming that the adversary cannot query a hash function
during the memory-challenge phase. We are not aware how this could be enforced
in practice.

A key security assumption made by all software-based memory-erasure protocols up
to 2019 is that prover and verifier should run the protocol in isolation, i.e.
without interference from an external attacker. This assumption is cumbersome to
enforce successfully for wireless communication, limiting the use of software-based
memory erasure to rather specific use-cases. Trujillo-Rasua questioned
in~\cite{Trujillo-Rasua2019} whether such assumption is necessary, offering a
symbolic security proof of a memory-erasure protocol that uses distance bounding
to thwart collusion between the prover and an external conspirator. 
This work,
however, left open the question of how to bound the probability of success of an
adversary (i.e.\ malware colluding with an external conspirator) passing the
memory-erasure test without removing the malware.  

In 2024, Bursuc et al.~\cite{Bursuc2024,Bursuc2024a} introduced the first
memory-erasure protocols with provable security bounds that do not depend on the
isolation assumption.   
Two of the protocols follow the methodology established
in~\cite{Dziembowski2011,Karvelas2014}, consisting of the construction of a
graph-based labelling function and an analysis of the protocol's security via a
reduction to a graph-pebbling game. The third protocol is an extension of Perito
and Tsudik's protocol with a distance-bounding technique. The goal is to
guarantee that the prover does not receive external help during the execution of
the protocol. This protocol is proven secure unconditionally, while those based
on graph-based labelling functions are proven secure within the random oracle
model. Like in~\cite{Trujillo-Rasua2019}, the three protocols introduced
in~\cite{Bursuc2024,Bursuc2024a} use a distance-bounding mechanism consisting of
several round-trip-time measurements, each requiring a round of communication
between prover and verifier. The number of round-trip-time measurements thus
become a security parameter in these protocols, which we denote \(r\) in
\Cref{tab:secprop}.

The SPEED protocol~\cite{Ammar2018} performs memory erasure using a Trusted
Software Module (TSM) as a lightweight alternative to hardware security modules.
In practice, it might add a significant overhead to any program running on the
platform, as the software-based memory protection continuously monitors the
platform to ensure memory isolation at all times. Since memory erasure can start
from any state of the device, it is more desirable to avoid imposing constraints
on programs running on it. SPEED uses distance-bounding to ensure that only a
nearby verifier can start the memory-erasure procedure. This does not make the
protocol secure in the presence of external attackers, unless the security of
the TSM guarantees that the prover device can only communicate with the verifier
during the run of SPEED.\@

We summarize our study of the literature on software-based memory-erasure
protocols in \Cref{tab:secprop}. The table shows to what extent each of the
protocols we have discussed satisfies the following relevant features, as
claimed in the literature. 

\begin{itemize}
    \item \textbf{Proof:} There exists a formal proof of security, which gives
    security guarantees with mathematical rigour.  
    \item \textbf{Prob.:} There exists a formula to bound the probability of
    success of the attacker given the device characteristics and protocol
    parameters. Unlike asymptotic analysis, this feature allows the analyst to
    measure the actual security of the protocol on a given device.  
    \item \textbf{No-Isolation:} The protocol does not assume that the device is
    isolated during the execution of the protocol, meaning that the protocol
    resists network attacks to some extent.
    \item \textbf{Erasure:} The proportion of the device's memory that can be
    erased. Together with security guarantees, this is the most important
    feature of a memory-erasure protocol.  
    \item \textbf{Time:} The time complexity of running  the protocol.
    \item \textbf{Comm.:} The communication complexity of running the protocol.
\end{itemize}

\securityprop{}

From \Cref{tab:secprop} one can already draw a preliminary and theoretical
comparison of software-based memory-erasure protocols. The series of protocols
\poselight{}, \posegraph{} and \poserandom{} are the only ones that come with a
formal security proof, a bound on the probability of success of the attacker,
and a mechanism to operate without relying on the isolation assumption. Their
erasure guarantee, however, is not asymptotically close to \( 1 \) for realistic
values of \( r \). Further, it is unclear how the parameter \(r\) affects their
computational and communicational complexity in practice. The other two
protocols that resist network attacks (no-isolation) are \textbf{TR} and
\textbf{SPEED}. The former provides no erasure guarantees, though, while the
latter comes with no security proof and has the drawback of relying on a device-specific Trusted Software
Platform that negatively impacts the performance of all programs running on the
device. The \dziembowski{} protocol does ensure full memory erasure while, at
the same time, it comes with a formal security proof. However, it offers the
worst computational complexity amongst all protocols, hinting that it might be a
viable option only for devices with low memory size. This computational
complexity is improved upon by \karvelas{}, but at the cost of erasing only a
small fraction of the device's memory. Lastly, \karame{} and \perito{} offer
optimal computational complexity, but offer neither a formal proof of security
nor a bound on the adversary's success probability.

To enable a more
detailed and accurate comparison of the protocols above, it is necessary to obtain empirical data on their
performance (last two columns of \Cref{tab:secprop}).  
That is the goal of the remainder of this article.

\section{\uppercase{Experimental Setup}}\label{sec-setup}

In this section we describe the procedure we have followed to implement and test
the protocols of interest, providing all the necessary details for the
reproducibility of the experiments.

\subsection{IoT proving devices, verifier device and communication channel}

For our experiments we considered 3 IoT microcontrollers produced by Texas
Instruments with different characteristics: FR5994, F5529, CC2652. These devices
acted as provers in the memory-erasure protocol. All prover implementations were
done in Portable C, making them usable on any platform or architecture. To
create the binaries, we configured the compiler to optimize for code size. A
personal Dell laptop acted as verifier, running Python 3 implementations of each
protocol. The Bluetooth protocol was used as communication channel between the
verifier and the prover.


\Cref{tab:devprop} depicts the characteristics of each device, namely memory
(code size~\footnote{Code size refers to the amount of memory occupied by the
executable code running on an IoT device. This includes, for example,
applications and libraries.} + data size~\footnote{Data size refers to the memory
used by the data used during execution. This includes, for example, variables
and buffers used by the running code.}), 
maximum clock speed, on-device crypto accelerators, Bluetooth version,
architecture, microcontroller unit family (MCU), and IoT class~\cite{rfc7228}.
Notice that some devices have hardware accelerators. For these, we also
experimented using a hardware-accelerated version of the hash function.

\devprop{}

\subsection{Hash function implementations}

Most software-based memory-erasure protocols invoke multiple hash
function calls,
while at the same time being independent of the particular hash
function that was implemented.
Hence, given their prominent role, we provide different hash function implementations to compare their
performance and memory requirements. 

We selected these functions according to popularity, amenability to
software-based implementations, and applicability to the IoT domain. All of them
calculate a digest of 256 bits. The following hash functions were selected:
\begin{itemize}
    \item ascon~\cite{Dobraunig2021}: a sponge-based hash function
    selected in the Lightweight Cryptography Standardization Process by
    NIST\footnote{\url{https://csrc.nist.gov/News/2023/lightweight-cryptography-nist-selects-ascon}}.
    \item blake2~\cite{Aumasson2013}: a widely deployed and highly efficient hash
    function. It was designed to be especially performant in software implementations.
    \item
    blake3\footnote{\url{https://github.com/BLAKE3-team/BLAKE3-specs/blob/master/blake3.pdf}}:
    a recent improvement on the blake2 hash function which is claimed to be much
    faster while offering similar security guarantees. This is the fastest (in
    software implementations) cryptographic hash function we could find.
    \item sha256~\cite{Eastlake2006}: a well known and widely used hash function
    based on the Merkle-Damgård construction proposed by NIST.\@ Up to this day,
    it is still considered secure, although it is prone to length extension
    attacks~\cite{Tsudik1992}.
    \item
    aeshash\footnote{\url{https://csrc.nist.rip/groups/ST/toolkit/BCM/documents/proposedmodes/aes-hash/aeshash.pdf}}:
    an unpublished hash function whose core utilizes AES instructions. It was
    selected mainly to check how much speed-up could be achieved in a device
    with only an AES accelerator (FR5994). 
    As far as we are aware, this hash function has not been thoroughly analysed,
    so it does not offer (yet) strong security guarantees, and it is only used
    here for the purpose mentioned before.
\end{itemize}

\Cref{tab:memHash} shows the memory footprint (in bytes) of the hash
implementations on each device. The lower the memory footprint the better, as
these bytes cannot be erased during the execution of the protocol. One
surprising result is that the memory footprint of the sha256hw function, which
uses the hardware accelerator, actually occupies more space than the
software-based implementation sha256. We conjecture this must be the result of
the extra code necessary to access the hardware module. In the case of the hash
implementation used by \perito{}, which originally uses an HMAC instantiated
with sha256, the memory occupied by the hash function is not shown in the table,
because the implementation used was tightly coupled with the HMAC routine,
making it incomparable to the others.

	\begin{table*}[!ht]
		\small
		\centering
	\begin{tabular}{lrrrrrr}
 & aeshash & ascon & blake2 & blake3 & sha256 & sha256hw \\
CC2652 & 1250 & 5869 & 6365 & 21185 & 1042 & 1398 \\
F5529 & --- & 2117 & 13997 & 22796 & 2254 & --- \\
FR5994 & 634 & 2117 & 13997 & 24694 & 2254 & --- \\
\end{tabular}

	\caption{\label{tab:memHash}Memory footprint (in bytes) of the hash functions on each
	device}
\end{table*}
{}

The memory footprint of each hash function varies according to the device, due
to the use of C implementations optimized for each architecture. For
example, the blake2 implementation used for the CC2652 is optimized for the 32
bit architecture, which is not suitable for the others. For some hash functions
this leads to lower sizes (blake2, blake3 and sha256) in the CC2652 with a 32bit
architecture, or lower sizes (the rest) in the F5529 and FR5994 with a 16bit
architecture. Taking into account the three devices, sha256 is the hash function
with the lowest memory footprint, followed by ascon and aeshash.

For reproducibility purposes, we remark that, to measure the memory footprint of
the hash function implementations on each device, we used as baseline a trivial
hash function with minimal memory footprint. This allows us to measure the
memory footprint of a hash function on a device by comparing it against the
implementation of the trivial hash function on the same device. We measure
memory in this way to ensure \first{} that the compiler code optimization
routine does not remove the hash function and \second{} that all hash functions
implementations share the rest of the program code.

\subsection{Protocol selection}

Out of the \(9\) protocols described in \Cref{sec-survey}, we selected 7
protocols for their implementation and evaluation. SPEED~\cite{Ammar2018} was
discarded because it relies on a memory-isolation technique that is neither open-source nor fully specified in the original article. 
We also
discarded the protocol in~\cite{Trujillo-Rasua2019} because the protocol
specification is given symbolically, abstracting away from important
implementation details, such as the number of round-trip-time measurements, the
size of the nonces, etc. Even though the author in~\cite{Trujillo-Rasua2019}
offers an instantiation of the symbolic protocol specification, it leaves as
future work the analysis of its security and erasure guarantees. 

We note that the protocols \poselight{}, \posegraph{} and \poserandom{}
\cite{Bursuc2024,Bursuc2024a}, depend on an additional security parameter \(r\)
that establishes the number of round-trip-time measurements performed by the
protocol. For our empirical setting, we set \( r = 71 \), which gives an erasure
guarantee of \( 90\% \) of the device's memory with probability \( (1 - 10^{-3})
\). If one wishes to erase \( 1/32 \) of the memory only, like in the
\karvelas{} protocol, then it would be sufficient to set \( r = 3 \), which is
efficient. If one wishes to erase \( 99\% \) of the device's memory, like in
\dziembowski{}, \karame{} and \perito{}, then the protocols
in~\cite{Bursuc2024,Bursuc2024a} would need to execute hundreds of
round-trip-time measurements, which is inefficient. We thus believe that \( r =
71 \) strikes a good balance between erasure guarantees and performance. This
means that, for the remainder of this article, when we refer to \posegraph{},
\poselight{} and \poserandom{}, we are assuming that they all execute \( 71 \)
round-trip-time measurements.

In \Cref{tab:memAlgo}, we show the memory footprint of our protocol
implementations for each combination of protocol and device. To calculate the
memory footprint of a protocol on a device, we considered a dummy implementation
of each protocol, that, while sending messages of the same size and in the same
order, occupies a negligible amount of memory. The rest of the code remained
exactly the same. Comparing this dummy implementation with the original one
allowed us to compute the actual size of each protocol implementation taking
into account compiler optimizations.

	\begin{table*}[!ht]
		\footnotesize
		\centering
	\begin{tabular}{lrrrrrrr}
 & \dziembowski{} & \karame{} & \karvelas{} & \perito{} & \posegraph{} & \poselight{} & \poserandom{} \\
CC2652 & 620 & 1023 & 1343 & 274 & 1182 & 1181 & 312 \\
F5529 & 777 & 1343 & 1868 & 356 & 1610 & 1644 & 415 \\
FR5994 & 781 & 1353 & 1876 & 357 & 1610 & 1644 & 416 \\
\end{tabular}

	\caption{\label{tab:memAlgo}Memory footprint (in bytes) of the protocol
	implementations on each device}
\end{table*}
{} 

Notice that the values in \Cref{tab:memAlgo} vary between devices because their
architectures are different. Another reason for the size variation is that we
used the best available implementation for each hash function and architecture.
In general, the CC2652 implementations take more space than the F5529 or FR5994
ones. This is probably due to the use of byte size variables throughout the
implementations, which can be more efficiently represented in the 16 bits
architecture than in the 32 bits architecture. 
Taking into account the three
devices, the protocols with smaller memory footprint are \perito{} and
\poserandom{}.

\subsection{Memory to be erased}

To be able to run all protocols in a reasonable time and avoid the need for
device-specific engineering, while erasing the same amount of memory with each
protocol, we set up our experiments in such a way that a fixed portion of the
memory of each device is erased. Our implementation creates an array with a
fixed size at compile time, and this array is exactly what is erased while
running the protocol. This makes our implementation simple and
device-independent. 
Because the array must fit in the data size of each device, we restrict ourselves to the following memory sizes. 
For the devices F5529 and FR5994 we erase exactly \(2\)KB, and for CC2652 we erase exactly 2 KB, 4 KB and 8 KB.\@
Increasing the maximum erased size in each device beyond the limits
mentioned before, led to the risk of overwriting memory addresses used
during execution, therefore making it unfeasible. 

We notice that
our implementations are limited to performance testing, and will need
to be adapted for deployment in a real setting. The reason being that, in practice, memory erasure protocols define the segment of memory to be erased prior compilation, rather than letting the compiler decides. 
Doing so, however,   
requires device-dependent
implementations, which we considered 
would add unnecessary complexity to the comparison task. 


\subsection{Distance}

The distance between the device (prover) and the laptop
(verifier) was approximately 1 meter. Each run of the protocol was done
independently,
as the main objective was to compare the protocols against each
other in the simplest possible setting.


\section{\uppercase{Experimental results}}\label{sec-results}

This section reports on the time each combination of protocol and hash
function implementation takes to \first{} erase a fixed amount of memory on a device and \second{} to generate a proof of erasure.   

\subsection{Impact of the hash function implementations}

Most protocols under analysis, namely \( \dziembowski, \karame, \karvelas, \posegraph \) and \( \poselight \), resort to several hash function calls to fill the device's memory with fresh data. This contrasts with the approach followed by \( \perito \) and \( \poserandom \) where the device's memory is filled with random data sent by the verifier. Hence, we start measuring the impact of the hash function implementation on the execution time of the former class of  protocols. 
We do so by measuring the execution time of the protocols right until
the point where the device's memory has been erased, thereby ignoring
the verification of the proof of erasure. We will refer to this (partial) execution time by \emph{erasure time}, to distinguish it from the \emph{verification time} where the verifier checks the erasure proof, and from the \emph{total execution time} which accounts for both: the erasure and verification time.     

\Cref{tab:computeTimeDevCCsVIII,tab:computeTimeDevFRsVI,tab:computeTimeDevFsVI}
display erasure time values for every combination of protocol and hash function, each table focusing on a given device and memory size to be erased. Notice
that the protocols \poserandom{} and \perito{} do not appear in the tables, as they do not call the hash function to fill the device's memory. 



	\begin{table*}[!ht]
		\centering
		\small
	\begin{tabular}{lrrrrrr}
 & aeshash & ascon & blake2 & blake3 & sha256 & sha256hw \\
\dziembowski{} & 10.9 & 31.5 & 5.1 & 3.8 & 7.3 & 1.7 \\
\karame{} & 0.0 & 0.0 & 0.0 & 0.0 & 0.0 & 0.0 \\
\karvelas{} & 7.4 & 20.9 & 3.6 & 2.7 & 5.1 & 1.2 \\
\posegraph{} & 14.8 & 60.0 & 5.9 & 4.4 & 8.7 & 2.1 \\
\poselight{} & 4.4 & 20.0 & 2.0 & 1.3 & 2.6 & 0.6 \\
\end{tabular}

	\caption{\label{tab:computeTimeDevCCsVIII}Erasure time in seconds on
	device CC2652 while attempting to erase \(8\)KB.}
\end{table*}
{}

	\begin{table}[!ht]
		\footnotesize
		\centering
	\begin{tabular}{lrrrrr}
 & aeshash & ascon & blake2 & blake3 & sha256 \\
\dziembowski{} & 3.1 & 80.5 & 20.1 & 15.6 & 27.8 \\
\karame{} & 0.1 & 0.3 & 0.1 & 0.1 & 0.2 \\
\karvelas{} & 4.9 & 130.8 & 34.5 & 26.9 & 48.1 \\
\posegraph{} & 8.6 & 218.7 & 56.7 & 44.0 & 78.5 \\
\poselight{} & 4.6 & 108.7 & 28.7 & 22.3 & 39.5 \\
\end{tabular}

	\caption{\label{tab:computeTimeDevFRsVI}Erasure time in seconds on
	device FR5994 while attempting to erase \(2\)KB.}
\end{table}
{}

	\begin{table}[!ht]
		\footnotesize
		\centering
	\begin{tabular}{lrrrr}
 & ascon & blake2 & blake3 & sha256 \\
\dziembowski{} & 47.5 & 12.6 & 9.9 & 17.4 \\
\karame{} & 0.7 & 0.6 & 0.6 & 0.6 \\
\karvelas{} & 76.3 & 21.2 & 16.5 & 30.1 \\
\posegraph{} & 128.9 & 34.0 & 26.3 & 48.2 \\
\poselight{} & 64.1 & 17.5 & 13.6 & 24.6 \\
\end{tabular}

	\caption{\label{tab:computeTimeDevFsVI}Erasure time in seconds on device
	F5529 while attempting to erase \(2\)KB.}
\end{table}
{} 

We observe that it is notably faster to erase memory when the
device has a higher clock frequency, as can be seen when comparing performance
in the CC2652 with respect to the FF5529 and the FR5994 devices. Amongst the hash
functions, sha265hw was the fastest in the CC2652 device, which was expected as it uses the hardware
accelerator. One surprising result is that ascon,
which is meant to be used in lightweight cryptography, had a consistently worst
performance. In every device blake3 was faster than blake2, although the
difference is around 25 percent only. Aeshash was much faster than other hash
functions in the FR5994 device, as expected by the usage of the AES hardware
module present on this device. On the contrary, it performed worse than
blake2, blake3 and even sha256 in the CC2652 device. Therefore, we deduce that
if the device is fast enough, pure software implementations might be more
performant than hybrid implementations such as the one used in aeshash.

To conclude, the impact of the hash function implementation is high in most of the protocols analysed, with the exceptions of \( \perito \) and \( \poserandom \). 
Interestingly, the smaller erasure time on each device is achieved with a different hash function. This suggests that, before deploying a memory-erasure protocol on a given device, it is worth testing it with different hash functions implementations. 
Lastly, cross-checking \Cref{tab:memHash} (memory cost) with \Cref{tab:computeTimeDevCCsVIII,tab:computeTimeDevFRsVI,tab:computeTimeDevFsVI} (computational cost), we conclude that \first{} aeshash is the fastest and smaller hash function implementation on device FR5994, \second{} sha256 and sha256hw outperform the other hash functions on the device CC2652, and \third{} on the device F5529 there is a clear looser, namely ascon, but no clear winner.

\subsection{Total execution time}

Next, we measure the total execution time of each combination of protocol and hash function implementation. This is admittedly the most important performance variable in our experiments. 
The values can be found in 
\Cref{tab:totalTimeDevCCsVIII,tab:totalTimeDevFRsVI,tab:totalTimeDevFsVI}, each table focusing on a given device and memory size to be erased.

	\begin{table*}[!ht]
		\footnotesize
		\centering
	\begin{tabular}{lrrrrrrr}
 & aeshash & ascon & blake2 & blake3 & none & sha256 & sha256hw \\
\dziembowski{} & 11.1 & 31.7 & 5.3 & 4.0 & --- & 7.5 & 1.9 \\
\karame{} & 23.4 & 23.4 & 23.3 & 23.3 & --- & 23.4 & 23.3 \\
\karvelas{} & 7.6 & 21.1 & 3.8 & 2.9 & --- & 5.3 & 1.3 \\
\perito{} & --- & --- & --- & --- & --- & 23.2 & --- \\
\posegraph{} & 21.4 & 66.6 & 12.5 & 11.0 & --- & 15.3 & 8.7 \\
\poselight{} & 11.0 & 26.6 & 8.6 & 8.0 & --- & 9.2 & 7.2 \\
\poserandom{} & --- & --- & --- & --- & 29.7 & --- & --- \\
\end{tabular}

	\caption{\label{tab:totalTimeDevCCsVIII}Total execution time in seconds on
	device CC2652 while attempting to erase  \(8\)KB.}
\end{table*}
{}

	\footnotesize
	\begin{table}[!ht]
		\footnotesize
		\centering
	\begin{tabular}{lrrrrrr}
 & aeshash & ascon & blake2 & blake3 & sha256 \\
\dziembowski{} & 3.2 & 80.7 & 20.2 & 15.7 &  27.9 \\
\karame{} & 2.4 & 2.6 & 2.5 & 2.4 & - 2.5 \\
\karvelas{} & 5.1 & 131.0 & 34.6 & 27.1 &  48.2 \\
\perito{} & --- & --- & --- & --- &  2.4 \\
\posegraph{} & 13.3 & 223.4 & 61.4 & 48.7 & 83.4 \\
\poselight{} & 9.3 & 113.3 & 33.5 & 27.2 &  44.4 \\
\end{tabular}

	\caption{\label{tab:totalTimeDevFRsVI}Total execution time in seconds on
	device FR5994 while attempting to erase \(2\)KB. For \poserandom{} it was
	6.8 seconds}
\end{table}
{}

	\begin{table}[!ht]
		\footnotesize
		\centering
	\begin{tabular}{lrrrrr}
 & ascon & blake2 & blake3 & sha256 \\
\dziembowski{} & 47.6 & 12.7 & 10.0 & 17.5 \\
\karame{} & 3.0 & 2.9 & 2.9 & 2.9 \\
\karvelas{} & 76.4 & 21.3 & 16.7 & 30.2 \\
\perito{} & --- & --- & --- & 2.9 \\
\posegraph{} & 133.5 & 38.7 & 31.1 & 52.9 \\
\poselight{} & 68.9 & 22.1 & 18.3 & 29.3 \\
\end{tabular}

	\caption{\label{tab:totalTimeDevFsVI}Total execution time in seconds on
	device F5529 while attempting to erase \(2\)KB. For \poserandom{} it was
	6.8 seconds}
\end{table}
{} 



In the CC2652
device, the fastest protocols were \karvelas{} and \dziembowski{}, despite the latter being the one
with worst (asymptotic) time complexity. We believe this to be possible because \first{} the constants hidden
in the asymptotic notation are very small for this protocol and somewhat larger
for the others, and \second{} the size of the erased memory is small. 
Looking at the protocols \perito{} and \poserandom, which follow the approach of sending a large random nonce over the network, we observe that they perform poorly in comparison to the others. Interestingly, that is not case if we shift our attention to the devices FR5994 and F5529. In those devices, 
\perito{} and \poserandom{} are amongst the fastest protocols. This difference in result can be explained by
the version of the Bluetooth protocol used. The CC2652 device has a Bluetooth
module included, able to run version 5.2, which means that in each challenge-response
round the message needs to go through the whole Bluetooth stack. On the other
hand, for the FR5994 and F5529 devices, a simple HC-05 module was used, which
makes the Bluetooth protocol overhead much smaller, as it does not support as
many features.

A summary of the results just described is given in \Cref{fig:totalTimeDevsVI}.
In that figure, we display the total execution time of each protocol on each
device by choosing the hash function implementation that gives the fastest
execution time. From the figure we derive that the choice of the most performant
(fastest) protocol for a device depends on the specific conditions in which it
will operate. In particular, clock speed, network cost and memory size determine
which protocol is more suitable. Cross-checking \Cref{tab:memAlgo} (memory cost)
with \Cref{fig:totalTimeDevsVI} (computational cost), we conclude that \perito{}
offers an optimal trade-off between memory footprint and execution time on the
devices F5529 and FR5994; other good alternatives are \dziembowski{} and
\poserandom{}. In the CC2652 the winner is \dziembowski{}.  

\totalTimeDevsVI{}

\subsection{How much does the size of memory impact the execution time?}

A rather surprising result from the empirical data we have presented so far is that the \dziembowski{} protocol seems to perform very well across all devices, despite having the worst asymptotic computational complexity. As the memory to be erased increases, one should expect a worse performance of 
\dziembowski{} in comparison to the other protocols. That is precisely
what we test next, i.e.\ how the execution time changes as we increase
the memory size. 
We perform this experiment on the CC2652 device, which allows us to erase up to 8KB of memory. \Cref{fig:totalTimeDevCCs} displays the results. 

\totalTimeDevCCs{}

Observe that the execution time of most protocols seem to increase linearly with the memory size. Exceptions are \posegraph{} and \poselight{}, whose execution time is dominated by the communication time rather than by the erasure time. The performance of \dziembowski{} on the erasure of 8KB of memory starts getting worse than that of \karvelas{}, supporting the hypothesis that its quadratic computational complexity might make it unsuitable for larger memory sizes.  
We acknowledge, that further experiments are needed to determine the memory threshold where \dziembowski{} starts behaving worse than the other protocols.

\section{\uppercase{Comparative analysis}}\label{sec-discussion}

Our results have not revealed a clear winner with respect to performance. That
said, \perito{}, \karame{}, \karvelas{} and \dziembowski{} seem to perform
relatively well across all devices.\ \karvelas{}, however, provides little
erasure guarantees (only \( 1/32 \) of the memory is guaranteed to be erased),
while \perito{} and \karame{} come with no formal security proof. Hence, our
next step is to provide a holistic comparison of the protocols, one where
security and erasure guarantees are considered alongside performance.  

\resultsummary{}

To reach our conclusions, we considered the following facts:

\begin{itemize}
    \item Protocols \karame{}, \karvelas{} and \perito{} offer a low level of
    security
    \item Protocols \poserandom{}, \posegraph{}, \poselight{} and \dziembowski{}
    offer a high level of security. Note, though, that the \dziembowski{} protocol is less secure
    than the others given that it is not resistant against distant attackers.
    \item Protocols \poserandom{}, \perito{} or \karame{} require sending the
    full memory of the device through the network.
    \item \dziembowski{} has quadratic complexity, hence its performance is
    worse when the amount of memory is large.
\end{itemize}

Next, we provide a fine-grained analysis of the results, by projecting them onto
specific use-cases. These results are summarized in \Cref{tab:resultsummary}.
They were obtained by extrapolating the behaviour of the protocols in each
device. 

\begin{itemize}
\item When the network cost is high, communication needs to be minimized,
therefore protocols such as \poserandom{}, \perito{} or \karame{} are impractical.
\begin{itemize}
    \item When memory is small, protocol \dziembowski{} is the clear winner (see
    for example \Cref{tab:totalTimeDevFsVI}). Even though it has quadratic complexity,
    its very low constant factor makes it faster than the rest of the protocols.
    We are considering that this protocol has high security even though it
    assumes the isolation assumption. If security against distant attackers is
    necessary, the winner for this category is the \poselight{} protocol.
    \item When memory is large, the most performant protocols are \poselight{}
    and \karvelas{}, for high security and low security, respectively.
\end{itemize}
\item When the network cost is low, the most performant protocols are
\poserandom{} and \perito{}, for high security and low security, respectively. A
surprising result in this setting is that \karame{} had a very similar
performance to \perito{}, which should not have been the case as it was
specifically designed to improve its performance. If the device in question has
hardware accelerators, then for the high security case it is possible that
\dziembowski{} is faster, as shown for example in
\Cref{tab:totalTimeDevCCsVIII}.
\end{itemize}

It is noteworthy that the clock speed only changed the selection of the winner
when the network cost is low and memory is small. In this scenario, the
\dziembowski{} protocol outperforms \poserandom{} and \karvelas{} outperforms
\perito{} (see for example \Cref{tab:totalTimeDevCCsVIII}).

\securityproptime{}

We end our analysis by completing the comparison table given in
\Cref{sec-survey} with the empirical data obtained during our experiments. This
provides another angle to compare the results obtained across devices, this time
contrasting performance with the same protocol features shown before in
\Cref{tab:secprop}. The results are displayed in \Cref{tab:secproptime}. For a
simpler visual comparison amongst the protocols, we labelled each cell in the
table with a symbol that indicates whether the protocol performs well (\cmark),
average (\qmark) or poorly (\xmark),   
 relative to the other protocols. For the numerical values in the table, we
 determined their labels by clustering the values in a column in three clusters.
 Specifically, we used \( k \)-means for clustering. For example, in the last
 column of the table, which gives the execution time of each protocol on the
 CC2652 device when erasing \( 8 \)KB, there are three clusters that minimize
 the sum of the squared Euclidean distances of each point to its closest
 centroid, namely \( \text{\cmark} = \{1.9, 1.3\} \), \( \text{\qmark} = \{8.7,
 7.2\} \) and \( \text{\xmark} = \{23.3, 23.2, 29.7\} \).   
The table reinforces the analysis results we have mentioned earlier. An
interesting observation is that  
\poselight{} is the only protocol that does not perform poorly on any of the
features considered.

\section{\uppercase{Conclusions}}\label{sec-conclusion}

In this paper we presented the outcome of our experiments with various
memory-erasure protocols in an IoT setting. We
implemented\footnote{\url{https://gitlab.com/uniluxembourg/fstm/dcs/satoss/memory-erasure-experiments}}
7 protocols, each with several variants depending on the hash function used, and
tested them on 3 modern IoT devices. Furthermore, we compared the security
guarantees provided by each protocol, and contrasted them with their performance
in a practical setting. 

Our results revealed that current memory-erasure protocols are practical,
although erasing the full memory securely could take several minutes for the
slower devices. Network speed might be faster than local computation, therefore
aiming at minimal communication complexity is not always the best
choice. For protocols that use hash functions, the choice of the
hash function may dramatically influence
the protocol performance and memory footprint. Finally, the most
performant protocol might not be the best according to the asymptotic complexity analysis,
as for small memory sizes the hidden constants may play a determining role.

\section*{Acknowledgements}

Reynaldo Gil-Pons was funded by the Luxembourg National
Research Fund, Luxembourg, under the grant AFR-PhD-14565947. 
Rolando Trujillo-Rasua was supported by a Ramón y Cajal grant (RYC2020-028954-I) from the Spanish Ministry of Science and Innovation and the EU, as well as by projects PROVTOPIA (PID2023-150098OB-I00), funded by MICIU/AEI/10.13039/501100011033 and FEDER (EU), and HERMES, funded by INCIBE and the EU's NextGenerationEU/PRTR.




\end{document}